\documentclass{ws-ijmpa}
\usepackage[super,compress]{cite}
\usepackage{graphicx}
\begin{document}
\title{Quantum Fluctuations of Fields and Stress Tensors}

\author{L. H. Ford}
\address{{Institute of Cosmology\\
 Department of Physics and Astronomy \\ 
Tufts University, Medford, Massachusetts 02155, USA} }

\maketitle

\begin{abstract}
This is a review of recent work on quantum fluctuations of the electric field and of stress tensor operators and their physical effects.
The probability distribution for vacuum fluctuations of the electric field is Gaussian, but that for quadratic operators, such as the
energy density, can have a more slowly decreasing tail, leading to an enhanced probability of large fluctuations. This effect is very sensitive
to the details of how the measurement is performed. Some possible physical effects of these  large fluctuations will be discussed.
\end{abstract}

\section{Introduction}
\label{sec:Intro}

Vacuum fluctuations of the quantized electric field are primarily responsible for the Lamb shift, and can also contribute to an increase in the
quantum tunneling of charged particle through a potential barrier. Both of these effects, as well as the Gaussian probability distribution of
the fluctuations, will be reviewed in Sec.~\ref{sec:linear}. The need to average the field in space or time with functions of compact support 
will also be discussed. Such functions are nonzero only in finite intervals, and are needed to describe physical measurements, which necessarily
occur in finite time intervals and spatial regions. Some important properties of these functions, especially their Fourier transforms, are summarized
in Sec.~\ref{sec:compact}. These Fourier transforms fall more slowly than an exponential function for large argument.

This slow rate of decrease has important implications for the probability distributions of quadratic operators, such as the energy density. These operators
must be averaged in time to have finite fluctuations. The resulting probability distributions are discussed in Sec.~\ref{sec:stress}, where it is argued that
these distributions can fall relatively slowly for large argument, specifically as an exponential of a fractional power. This makes large vacuum fluctuations 
more likely than one might have expected, and may lead to observable effects, as discussed in Sec.~\ref{sec:phy-effect}. Section~\ref{sec:falsevac} 
deals with false vacuum decay in field theory. The key results are
summarized in Sec.~\ref{sec:final}. Units in which $\hbar = c =1$ are used throughout.

\section{Linear Field Operators}
\label{sec:linear}

\subsection{Spacetime averaging and the variance}

In this section, we discuss the fluctuations of a free, linear quantum field such as the electric field. A simple example which contains many essential features is the 
time derivative of a massless scalar field, which is similar to a cartesian component of the electric field.
This operator may be expanded in terms of creation and annihilation operators as
\begin{equation}
\dot{\varphi}({\bf x},t) = \sum_{\bf k} (a_{\bf k} \, F_{\bf k} +  a^\dagger_{\bf k} \, F^*_{\bf k})\,.
\label{eq:dot-phi} 
\end{equation}
Here quantization in a finite volume $V$ with periodic boundary conditions is assumed. The mode functions may be taken to be
\begin{equation}
F_{\bf k} = \sqrt{\frac{\omega}{2 \, V}}\, {\rm e}^{i({\bf k} \cdot {\bf x} - \omega t)} \,,
\end{equation}
where $\omega = |{\bf k}|$.

We wish to consider a spacetime average of the local operator $\dot{\varphi}({\bf x},t)$, given by averaging with a temporal sampling function $f(t)$ and a spatial sampling
function $g({\bf x})$,. The averaged operator is
\begin{equation}
\bar{\dot{\varphi}} = \int dt \, f(t)\, \int d^3 \,g({\bf x})  \; \dot{\varphi}({\bf x},t) = \sum_{\bf k} \sqrt{\frac{\omega}{2 \, V}}\,  
 (a_{\bf k} +  a^\dagger_{\bf k}) \,\hat{f}(\omega) \, \hat{g}( {\bf k} )\,.
 \label{eq:bar-dot-phi}
\end{equation}
Here $\hat{f}(\omega)$ and $\hat{g}( {\bf k} )$ are Fourier transforms of the sampling functions, given by 
\begin{equation}
\hat{f}(\omega) = \int dt  f(t)\,  {\rm e}^{i \omega t}\,
\end{equation}
and 
\begin{equation}
\hat{g}( {\bf k} ) = \int d^3x\, g({\bf x})\, {\rm e}^{i{\bf k} \cdot {\bf x}}\,.
\end{equation}
We will take both sampling functions to be real and even, which implies that $\hat{f}(\omega)$ and $\hat{g}( {\bf k} )$ are also real and even. The sampling functions
are assumed to be non-negative and normalized so that
\begin{equation}
\int dt \, f(t)\, = \int d^3x \,g({\bf x}) = 1\,.
\end{equation}
This implies
\begin{equation}
\hat{f}(0) = \hat{g}(0) = 1\,.
\end{equation}

In the vacuum state, $\bar{\dot{\varphi}}$ undergoes fluctuations with a vanishing mean value, $\langle 0|\bar{\dot{\varphi}}|0\rangle =0$, and a variance of
 \begin{equation}
\sigma = \langle 0 |{\bar{\dot{\varphi}}}^2|0\rangle = \sum_{\bf k} \frac{\omega}{2 \, V}\, \hat{f}^2(\omega) \, \hat{g}^2( {\bf k} )  \,.
\end{equation}
 This variance is finite so long as either $\hat{f}$ or $\hat{g}$ decrease sufficiently rapidly for increasing argument, which will be the case of the functions we consider.
 Thus, averaging in time alone or in space alone is sufficient to render $\sigma$ finite. In the continuum limit, $V \rightarrow \infty$, we have
 \begin{equation}
\sigma  = \frac{1}{2(2\pi)^3} \int d^3k \, \omega \, \hat{f}^2(\omega) \, \hat{g}^2( {\bf k} ) \,.
\label{eq:var-linear}
\end{equation} 
Let $\tau$ be the temporal sampling scale, or the characteristic width of $f(t)$. Then $ \hat{f}(\omega) \rightarrow 0$ if $\omega \gg 1/\tau$. If $\ell$ is the spatial
sampling scale, then   $ \hat{g}( {\bf k} )  \rightarrow 0$ if $\omega =  |{\bf k}|\gg 1/\ell$. Thus if $\tau \gg \ell$, temporal sampling dominates and $\sigma \propto 1/\tau^4$.
Similarly, if $\ell \gg \tau$, spatial sampling dominates and $\sigma \propto 1/\ell^4$.

\subsection{Probability distribution}

Here we review the result that the fluctuations of a linear operator, such as $\bar{\dot{\varphi}}$, obey a Gaussian probability distribution. One way to see this is
through a calculation of the moments of $\bar{\dot{\varphi}}$. Let the $n$-th moment be defined by
\begin{equation}
\mu_n = \langle (\bar{\dot{\varphi}})^n \rangle \,,
\end{equation}
 where the expectation value is taken in the vacuum state. Here we take $n$ to be an even integer, as $\mu_n = 0$ for $n$ odd. We may calculate  $\mu_n$ 
 explicitly by use of Wick's theorem to find 
 \begin{equation}
\mu_n = (n-1)!! \; \sigma^{n/2}\,.
\label{eq:mu2-lin}
\end{equation}
Note that Wick's theorem is most commonly used to express a time ordered product of operators in terms of a normal ordered product and products of contractions,
which are just factors of $\sigma$. Time ordering is not relevant for time independent operators such as $\bar{\dot{\varphi}}$, and the vacuum expectation value
of the normal ordered product vanishes. Finally, the factor of $(n-1)!!$ is a combinatorial factor describing the number of ways of contracting $(\bar{\dot{\varphi}})^n$.
 The set of moments given in Eq.~(\ref{eq:mu2-lin}) corresponds to the Gaussian distribution
 \begin{equation}
P(x) = \sqrt{\frac{1}{ 2 \pi\, \sigma}}\, {\rm e}^{-x^2/(2 \sigma)}\, ,
\label{eq:gaussian}
\end{equation}
as may be seen by calculation of the moments directly from $P(x)$ by
\begin{equation}
\mu_n = \int_{-\infty}^\infty P(x) \, x^n \, dx \,.
\end{equation}

The final step in the argument comes from  the Hamburger moment theorem\cite{Simon}, which states that a probability distribution is uniquely determined by its moments if
there exist constants $C$ and $D$ such that
\begin{equation}
|\mu_n| \leq C \, D^n\, n! 
\label{eq:Hamburger}
\end{equation}
for all $n$, which is satisfied by the moments given in Eq.~(\ref{eq:mu2-lin}). Later in this review, we will encounter situations where the Hamburger criterion is
not satisfied. 

\subsection{Some physical examples}

\subsubsection{Vacuum electric field fluctuations}

Vacuum fluctuations of the quantized electric field operator can have observable consequences. One example is the Lamb shift in the hydrogen atom, an upward shift
of the $2S$ level relative to the $2P$ level by an energy corresponding to a frequency of 1046 MHz. These two levels are degenerate in relativistic quantum mechanics,
so the Lamb shift is a quantum field effect. The full quantum electrodynamics calculation is rather complex, but the primary contribution comes from electric field fluctuations,
as was shown by Welton~\cite{Welton}. Welton argued that the electric field fluctuations cause the electron to move slightly further from the nucleus, slightly increasing the energy.
Because the  $2S$ 
wavefunction is nonzero at the nucleus, it is affected more than is the $2P$ state, where the wavefunction vanishes at the nucleus. 

A second system where vacuum electric field fluctuations can produce a potentially observable effect is in quantum tunneling of a charged particle through a potential 
barrier~\cite{FZ99,Huang:2015lea}. Here the fluctuations can give the electron a small kick, the net effect of which is to increase the tunneling probability by a small 
fraction of the order of $1\%$. This effect arises in perturbative quantum electrodynamics from the vertex diagram, describing a radiative correction to the scattering
amplitude.

A related but possibly somewhat larger effect might occur in the Casimir effect, where the presence of reflecting boundaries can modify and enhance the  vacuum electric field 
fluctuations. This effect was recently discussed in Ref.~\citen{F22}, where it was suggested the resulting enhanced quantum tunneling rates might explain some
experimental results of  Moddel {\it et al}~\cite{Moddel1,Moddel2}. These authors found that the presence of a reflecting plate can increase the current flowing through
a metal-insulator-metal interface.

\subsubsection{Density perturbations in inflationary cosmology}
\label{sec:inflation}

A very different system where vacuum fluctuations of a linear quantum field could produce observable effects is in the early universe. Different versions of the inflationary
model were first proposed by Starobinsky~\cite{Starobinsky80} and by Guth.\cite{Guth81} This model often involves a scalar field, the inflaton, whose energy density 
and pressure drives a period of exponential expansion of the universe. For a recent review, see Vazquez~{\it et al}\cite{Vazquez20}. A remarkable prediction of  
inflationary cosmology is that quantum fluctuations of the inflaton field can produce the initial spectrum of density perturbations which later grow to form structure in
the universe, such as galaxies and clusters of galaxies~\cite{CM81,GP82,H82,S82,BST83,Brandenberger85}. In a typical version of scalar field driven inflation, the
inflaton  behaves as a nearly massless classical field $\varphi(t)$ slowly evolving (slow roll) in a nearly flat potential, $V(\varphi)$. When the magnitude 
of the field reaches a critical value, $\varphi_c$, the evolution becomes more rapid and standard model particles are created (reheating), leading to the end of
inflation and a transition to a radiation dominated universe. However, the inflaton field is subject to small quantum fluctuations, $\delta \varphi$,  around its mean value.
These quantum fluctuations lead to density perturbations in the following picture:\cite{GP82} Take $\varphi(t)$ to be slowly increasing. Then a local region in which
 $\delta \varphi > 0$ will reach $\varphi_c$ and hence reheat sooner than surrounding regions. This region will begin redshifting soon, and hence become under dense
 compared to its neighbors. In contrast, a region in which  $\delta \varphi < 0$ will become a local over density, and is likely later to form galaxies and clusters of
 galaxies by gravitational collapse. This picture of primordial density perturbations arising from vacuum fluctuations of a nearly linear quantum field makes two predictions:
 a nearly scale invariant spectrum of perturbations, meaning that the expected magnitude is approximately independent of length scale, and perturbations described by
 a Gaussian probability distribution. Both of these predictions seem to be supported by cosmological observations.

\section{Sampling Functions with Compact Support}
\label{sec:compact}

It is desirable that the sampling functions $f(t)$ and $g({\bf x})$ have compact support, meaning that they are strictly equal to zero outside of a finite region.
This restriction comes because the sampling functions are intended to model a physical measurement, which necessarily occurs in  finite space and
time intervals. This implies that the sampling functions cannot be analytic, but we will require that they be infinitely differentiable. Compactly supported test
functions are used in rigorous approaches to quantum field theory~\cite{PCT}, but as a formal device to treat operator valued distributions, and are not given any 
physical interpretation.

A compactly supported, infinitely differentiable function will have a Fourier transform that decreases faster than any power, but more slowly than an exponential
function. A class of such functions of time was treated in Sec.~II of Ref.~\citen{FF2015}. Here the Fourier transform decays as an exponential of
a fractional power:
\begin{equation}
\hat{f}(\omega) \sim {\rm e}^{-(\tau \omega)^\alpha} \quad \tau \omega \gg 1\,,
\label{eq:asy-Fourier}
\end{equation}
where $0 < \alpha < 1$. The case $\alpha = \frac{1}{2}$ has special interest, as there is an electric circuit in which the current switches on in accordance with
the $f(t)$ for this case. In general, if $f(t)$ switches on at $t=0$, its form near this point is
\begin{equation}
f(t) \sim D \, t^{-\mu}\, {\rm e}^{-w\, t^\nu}\,,
\end{equation}
where the constants $D$, $w$, $\mu$, and $\nu$ are functions of $\alpha$. In particular, $\nu = \alpha/(1-\alpha)$, so $\alpha = \frac{1}{2}$ corresponds to a
temporal switch on of the form $f(t) \sim {\rm e}^{-1/t}$ as $t \rightarrow 0^+$. This class of compactly supported functions are special cases of the 
Fox H-function~\cite{Fox}.

A crucial feature of compactly supported functions is the relatively slow decay of the Fourier transform for frequencies $\omega \gg \tau^{-1}$ illustrated in
Eq.~(\ref{eq:asy-Fourier}). This leads to relatively large contributions of high frequencies to physical quantities, such as the variance in Eq. ~(\ref{eq:var-linear}).

\section{Quantum Stress Tensor Fluctuations}
\label{sec:stress}

Our primary topic will be the probability distribution for components of the stress tensor for a quantized field, such as the energy density. Even before we begin a 
discussion of fluctuations, we need to ensure that the expectation value of this operator is well defined. On a curved background spacetime, this is difficult
problem involving regularization and renormalization of parameters in the Einstein equations, including Newton's constant and the cosmological constant. Here
we restrict our attention to Minkowski spacetime, where normal ordering with respect to the Minkowski vacuum state is sufficient. The amounts to setting the mean 
value of the operator, about which fluctuations in the vacuum state occur, to be zero. Off-diagonal components of the stress tensor, such as an energy flux $T_{xt}$,
require no normal ordering. By symmetry, such components can have either sign. As a result, the corresponding probability distribution is symmetric; a negative
value will arise with the same probability as a positive value with the same magnitude.

The case of operators which are classically non-negative, such  the energy density, is more subtle. It is well known that in quantum field theory there exist quantum
states in which the expectation value of energy density can be negative in some regions. However, the magnitude and duration of this negative energy density is constrained by
quantum inequalities. Let $\langle \rho(t) \rangle$ be the normal ordered energy density at one space point. The time average of this quantity obeys an inequality
of the form (See  Ref.~\citen{F2017} for a recent review.)
\begin{equation}
\int \langle \rho(t) \rangle\, f(t)\, dt \geq -\frac{C}{\tau^d} \,,
\label{eq:QI}
\end{equation}
where $\tau$ is the characteristic width of $f(t)$, $C$ is a dimensionless constant, and $d$ is the number of spacetime dimensions. The physical content of
this inequality is the following: a measurement on a short timescale can observe a negative energy density with a relatively large magnitude, but this magnitude
decreases as the observation time increases. If Eq.~(\ref{eq:QI}) is the optimal bound, then there will exist some quantum state for which the inequality becomes
an equality.

We now turn to the probability distribution for the averaged energy density operator, $\bar{\rho} = \int  \rho(t) \, f(t)\, dt$. This distribution must have a lower bound
at the right hand side of Eq.~(\ref{eq:QI}), if the bound is optimal,
 because this is the lowest eigenvalue  of $\bar{\rho}$. That is, smallest value that can be found in a measurement is
this eigenvalue, which is also the smallest possible expectation value of the operator in any state. This is the optimal quantum inequality bound. Because this bound 
is negative, there are eigenstates of $\bar{\rho}$ with negative eigenvalues and a measurement of the averaged energy density has a nonzero probability of
yielding a negative outcome. Hence $P(x) \not= 0$ for values of $x$ larger than the optimal quantum inequality bound. 

\subsection{A Two Dimensional Example}
\label{sec:2D}

\begin{figure}[b]
\centerline{\includegraphics[width=10cm]{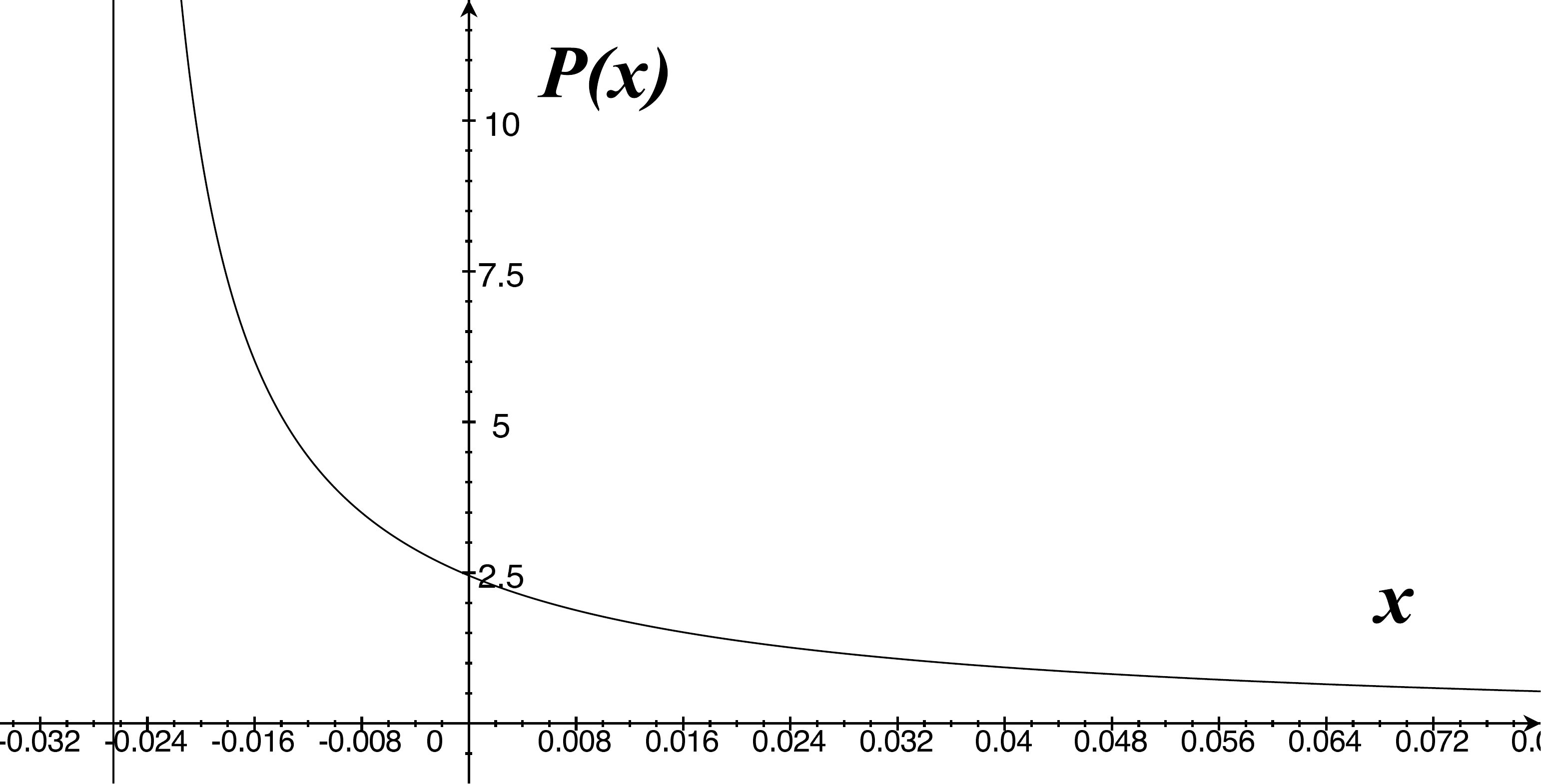}}
\caption{The probability distribution for the time averaged energy density of a massless scalar field in two dimensional Minkowski spacetime.
  The lower limit of $P(x)$ occurs at $x=x_0= -1/(12 \pi)$, illustrated by the vertical line.      
\label{fig:2D}}
\end{figure}

Here we consider the energy density of a massless scalar field in two dimensional spacetime, which was given in Ref.~\citen{FewsterFordRoman:2010}. 
Here $d = 2$ and we may let $x$ be an eigenvalue of the dimensionless energy density operator,  $\tau^2\, \bar{\rho}$. For a particular choice of $f(t)$, it is possible to
find the probability distribution, $P(x)$, explicitly as a gamma distribution function:
\begin{equation}
P(x) = \vartheta(x-x_0)\frac{\pi^{1/12}(x -x_0)^{-11/12}}{\Gamma(1/12)}\, 
{\rm e}^{-\pi(x-x_0)}\, . 
\label{eq:gamma-dist}
\end{equation}
Here $\Gamma$ is the gamma function and $\vartheta$ is a step function, which insures that $P(x) = 0$ for $x < x_0= -1/(12 \pi)$. This is the quantum inequality 
bound for this case, which was found by Flanagan~\cite{Flanagan97}, and shown to be the optimal bound. Thus $C = x_0$ in Eq.~(\ref{eq:QI}). This distribution
is illustrated in Fig.~\ref{fig:2D}\footnote{Originally published as Fig. 1 in Ref.~\citen{FewsterFordRoman:2010}.}, and satisfies
\begin{equation}
\int_{x_0}^\infty P(x) \, dx =1\,,
\label{eq:P-norm}
\end{equation}
as required of a probability distribution. 

Although the magnitude of negative outcomes in a measurement of $x$ is bounded, there is no upper limit on the values of $x$ for which $P(x)\not=0$, 
so arbitrarily large positive outcomes are possible.
However, the area under the $P(x)$ curve for $x<0$ is about $0.84$, so $84 \%$ of measurements will yield a negative value, and only $16\%$ a positive
value. The positive outcomes tend to be larger in magnitude, so the mean value is zero. Here $P(x)$ contains an integrable singularity at $x  = x_0$. Note that
although the probability distribution satisfies Eq.~(\ref{eq:P-norm}), it is possible to have $P(x) \gg 1$ in a small interval.

Further results in two dimensional spacetime are given in Refs.~\citen{Fe&Ho05,Fe&Ho18,AF19}.

\subsection{Results in Four Spacetime Dimensions}
\label{sec:4D}

Here we consider the quadratic operator defined as the spacetime average of the normal ordered square of  $\dot{\varphi}({\bf x},t)$
\begin{equation}
T =  \int dt \, f(t)\, \int d^3 \,g({\bf x})  \;   :(\dot{\varphi}({\bf x},t))^2: \,.
\label{eq:T-def}
\end{equation}
This operator arises in the components of the stress tensor for a massless scalar field, and is a convenient test case for the general behavior of these components,
such as the energy density. The mode expansion, Eq.~(\ref{eq:dot-phi}), leads to
 \begin{equation}
T = \sum_{i\, j} (A_{i j}\, a^\dagger_i \,a_j + B_{i j}\, a_i \,a_j + 
B_{i j} \, a^\dagger_i \,a^\dagger_j ) \,,
\label{eq:T}
\end{equation}
where 
\begin{equation}
A_{j \ell} = \frac{\sqrt{\omega_j \omega_\ell}}{V} \, \hat{f}(\omega_j -\omega_\ell )\, \hat{g}({\bf k}_j -{\bf k}_\ell ) 
\label{eq:A}
\end{equation}
and 
\begin{equation}
B_{j \ell} = \frac{\sqrt{\omega_j \omega_\ell}}{2 V} \, \hat{f}(\omega_j +\omega_\ell )\, \hat{g}({\bf k}_j +{\bf k}_\ell ) \,.
\label{eq:B}
\end{equation}

The moments are vacuum expectation values of powers of $T$  
\begin{equation}
\mu_n = \langle 0| T^n |0 \rangle \,,
\label{eq:mu-n}
\end{equation}
 and are expressible as $n$-th degree polynomials in the $A_{j \ell}$ and   $B_{j \ell}$. In particular, the second moment
or variance is $\mu_2 = 2\sum_{j \ell} B_{j \ell}^2$. In the $V \rightarrow \infty$ limit, it becomes
\begin{equation}
\mu_2 =   \frac{1}{2 (2\pi)^6}   \int d^3k \,  d^3k' \, \omega \,  \omega'\; \hat{f}^2(\omega+\omega' )\, \hat{g}^2({\bf k}+{\bf k'} ) \,.
\label{eq:mu2}
\end{equation}
In contrast to the case of a linear operator, the moments of quadratic operator are finite only with time averaging. With spatial averaging alone, the factors of $\hat{f}$ would not
appear in the above expression, and the integral would receive a divergent contribution from regions where ${\bf k} = -{\bf k'}$. For a similar reason, quantum  inequalities in
four dimensional spacetime require temporal averaging~\cite{FHR2002}.

\subsection{Probability Distributions}
\label{sec:pdf}

The treatment of the probability distribution for a quadratic operator in four dimensional spacetime,  such as $T$, is more complicated than in the case of linear operators. 
One approach is to examine the rate  of growth of the moments $\mu_n$ as $n$ increases~\cite{FFR2012,FF2015,FF2020}.
 
 \subsubsection{Moments Approach: Worldline Limit}
 \label{sec:moments-worldline}
 
 This was first done in Ref.~\citen{FFR2012} for the case of $\dot{\varphi}^2({\bf x},t)$ averaged in time only with a Lorentzian function. This is not a compactly supported
 function, but its Fourier transform is an exponential, and hence the $\alpha \rightarrow 1$ limit of the functions described by Eq.~(\ref{eq:asy-Fourier}). Here the special
 properties of a Lorentzian allowed the explicit computation of a finite set of moments ($n \leq 65$) by an algebraic computing program. The result is that the moments
 grow as 
 \begin{equation}
\mu_n \propto  (3n)!
\label{eq:mu-Lorentian}
\end{equation}
 for $n \gg 1$, a remarkably rapid growth, which implies that the asymptotic tail of the probability distribution must fall relatively slowly with increasing $x$.
 However, the Hamburger condition Eq.~(\ref{eq:Hamburger}) is clearly not satisfied here. There is a weaker condition for uniqueness which applies to a probability distribution
 which is nonzero on a half line, applicable to operators such as $\dot{\varphi}^2({\bf x},t)$ which are non-negative in classical physics. This is the Stieltjes condition,
 which is the same as Hamburger condition, but with $n!$ replaced by $(2 n)!$,
\begin{equation}
|\mu_n| \leq C \, D^n\, (2 n)! \,.
\label{eq:Stieltjes}
\end{equation}
However, this condition is also not fulfilled here, so the moments do not uniquely determine $P(x)$.

Nonetheless, we may attempt to infer some features of the asymptotic form of $P(x)$. Assume that
 \begin{equation}
P(x) \sim c_0 \, x^b\, {\rm e}^{-a x^c},  \, x \gg 1 \, ,
\label{eq: asym-P}
\end{equation}
for some constants $c_0$, $b$, $a$, and $c$. Because $P(x) = 0$ for $x < x_0$, the quantum inequality bound, the moments are
\begin{equation}
\mu_n = \int_{x_0}^\infty P(x) \, x^n \, dx \,.
\end{equation}
If we assume $n \gg 1$, we may use the postulated asymptotic form in Eq.~(\ref{eq: asym-P})  and ignore the interval $[x_0,0]$. Then
\begin{equation}
\mu_n   \sim c_0 \, \int_{0}^\infty  x^{n +b}\, {\rm e}^{-a x^c}\, dx = \frac{c_0}{c} \, a^{(n+b+1)/c} \, [(n+b+1)/c -1]! \, .
\label{eq: asym-mu}
\end{equation}
 The most important constant in Eq.~(\ref{eq: asym-P}) is the exponent $c$. The $(3n)!$ growth of the moments from Eqs.~(\ref{eq:mu-Lorentian}) leads to
  $c = 1/3$ in this case. This implies  a relatively high probability for large fluctuations, at least as compared to that predicted by a Gaussian distribution.
 The other constants in the asymptotic form Eq.~(\ref{eq: asym-P}) may also be determined, not only for $\dot{\varphi}^2({\bf x},t)$, but also for several
 related operators, including the scalar and electromagnetic energy densities and the squared electric field. The results are listed in Table IV of Ref.~\citen{FFR2012}.
 In all of the cases studied using a Lorentzian time sampling function,  $c = 1/3$, $b=-2$, but $c_0$ and $a$ vary slightly but are somewhat less than one.
 
 The issue of the non-uniqueness of the probability distribution remains to be addressed. Because the Stieltjes condition, Eq.~(\ref{eq:Stieltjes}) is not fulfilled,
 there can exist two distinct probability distribution functions, $P_1(x)$ and  $P_2(x)$, with exactly the same moments, which would imply that
 \begin{equation}
\int_{x_0}^\infty [P_1(x) - P_2(x)]    \, x^n \, dx  = 0
\end{equation}
for all $n$. This can only happen if the difference $P_1(x) - P_2(x)$ is an oscillatory function of $x$. However, in physical applications of the  probability distribution,
we are not interested in the values of $P(x)$ at a specific point, but rather integrals over finite ranges, which give the probability of an outcome in that range. For
example, one might be interested in the probability of an outcome where $x \geq y$, which is the probability of a fluctuation at least as large as that described by 
the value $y$. This probability is given by the complementary cumulative distribution
\begin{equation}
P_>(y) = \int_y^\infty P(x) \, dx\,,
\end{equation}
which can be expected to be relatively insensitive to oscillating contributions to  $P(x)$. We will return to the uniqueness issue later in Sec.~\ref{sec:moments-STA}.

The approach used in Ref.~\citen{FFR2012} is limited to the case of a Lorentzian sampling function. A more general method, which can apply to compactly supported
functions, was developed in Ref.~\citen{FF2015}. Here the moments cannot in general be computed explicitly, but some general features of their growth for large $n$
can be inferred from an iteration procedure. Consider  the case of  $\dot{\varphi}^2({\bf x},t)$ averaged in time with one of the functions discussed in Sec.~\ref{sec:compact},
whose Fourier transform has the asymptotic form given in Eq.~(\ref{eq:asy-Fourier}). The key result is that the moments grow as
\begin{equation}
\mu_n \sim (3 n/\alpha)! \,.
\label{eq:compact-time}
\end{equation}
In the limit that $\alpha \rightarrow 1$, this agrees with  Eq.~(\ref{eq:mu-Lorentian}).

The resulting probability distribution is of the form of Eq.~(\ref{eq: asym-P}) with $c = \alpha/3$. Thus, switching corresponding to $\alpha = 1/2$, as could be produced 
by the electrical circuit described in Sec.~IIC of  Ref.~\citen{FF2015}, will produce a probability distribution falling as an exponential of $x^{1/6}$.

\subsubsection{Moments Approach: Spacetime Averaged behavior}
 \label{sec:moments-STA}

The effects of spatial as well as temporal averaging on the rate of growth of the moments was treated in Ref.~\citen{FF2020}. Again an iteration procedure was developed
to find approximations to the $n$-th moment, given by Eq.~(\ref{eq:mu-n}) for $n \gg 1$.  Here we assume spatial sampling over a finite spherically symmetric region, so
$g({\bf x}) = g(|{\bf x}|)$ and $\hat{g}({\bf k}) = \hat{g}(k)$, where $k = |{\bf k}|$. We take the asymptotic form of $\hat{g}(k)$ to be of a  form analogous to Eq.~(\ref{eq:asy-Fourier})
\begin{equation}
\hat{g}(k) \sim {\rm e}^{-(\ell \, k)^\lambda} \quad \ell \, k \gg 1\,,
\label{eq:asy-Fourier-g}
\end{equation}
where $\ell$ is the characteristic spatial sampling scale and $0 < \lambda < 1$ is a constant.

If $\ell \ll \tau$, then the worldline approach described in Sec.~\ref{sec:moments-worldline} will be a good approximation for a finite range, specifically
$x <  x_*$, where
\begin{equation}
x_* \approx \left(\frac{\tau}{\ell}\right)^3\,.
\label{eq:x*}
\end{equation}
If the spatial scale is small compared to the temporal scale, so that $x_* \gg 1$, then there will be a finite range where $x \gg 1$, but $x <  x_*$. This is a worldline regime
where $P(x)$ has the form in Eq.~(\ref{eq: asym-P}) with $c \approx \alpha/3$. 

When $x \approx x_*$, there is a transition to the region of spacetime averaged behavior. If $\alpha \ge \lambda$, in this region $P(x)$ again has the form in Eq.~(\ref{eq: asym-P}) 
 if $x \gg 1$, but  with $c \approx \alpha$. The increase in the value of the exponent $c$ due to spatial averaging has the effect of decreasing the probability of 
 large fluctuations compared to the worldline behavior. It is of interest to note that in this regime, the moments are now  growing as $\mu_n \sim (n/\alpha)!$ for large $n$.
 This implies that if $\alpha \geq 1/2$, the Stieltjes criterion, Eq.~(\ref{eq:Stieltjes}) is fulfilled, and the probability distribution is uniquely determined by its moments. 

In summary, quadratic massless field operators in four dimensional spacetime, which have  dimensions of ${\rm length}^{-4}$, have the following general forms for their asymptotic
probability distribution $P(x)$: In the worldline limit, $x < x_*$, the distribution has the form of Eq.~(\ref{eq: asym-P}) with $c=\alpha/3$. Here we assume a compactly supported
sampling function whose Fourier transform decays according to Eq.~(\ref{eq:asy-Fourier}). In the spacetime averaged limit, $x >x_*$ with space as well as temporal sampling, the exponent
$c$ becomes $c \approx \alpha$. Thus for the $\alpha = 1/2$ case, there is a transition from $c = 1/6$ behavior to a region where $c = 1/2$. This applies to a wide class of operators,
including $\dot{\varphi}^2({\bf x},t)$, and the energy density for the massless scalar and electromagnetic fields. This class also includes the momentum density operator for
the electromagnetic field which will be discussed in Sec.~\ref{sec:RadPress-charges}.

\subsubsection{The Diagonalization Approach}
\label{sec:diag}

There is a completely different approach to finding the probability distribution, which was developed in Refs.~\citen{SFF18} and \citen{WSF21}. This involves diagonalization
of the quadratic operator in the form of Eq.~(\ref{eq:T}) by a Bogolubov transformation and construction of the operator's eigenstates and eigenvalues. In practice, this
requires the use of a finite number of modes and numerical computations. Here we outline the basic ideas, and then summarize the numerical results.

The creation and annihilation operators  for mode  $j$,  $a^\dagger_j$ and  $a_j$, respectively, which appear in Eq.~(\ref{eq:T}), correspond to physical particles, and the state
of the system is taken to be the vacuum state, for which $ a_j |0\rangle_a =0 $ for all $j$. We perform a Bogolubov transformation to a new set of creation and annihilation
operators,  $b^\dagger_j$ and  $b_j$. This is a linear transformation of the form
\begin{equation}
a_j=\sum\limits_k (\alpha _{jk} b_k-\beta _{jk} b_k^\dagger) \,.
                                      \label{eq:Bogo1}
\end{equation}
If we deal with a system with $N$ modes, then the Bogolubov coefficients,  $\alpha _{jk}$ and $\beta _{jk}$, are the components of a pair of $N \times N$ matrices.

The operator $T$ is diagonal in the new set  of creation and annihilation operators if it takes the form
\begin{equation}
T = \sum\limits_j \lambda_j\, b^\dagger_j b_j \: + C_{\text{shift}} \, \mathbb{I}\,,
\end{equation}
where $\mathbb{I}$ denotes the $N \times N$ identity matrix, and $C_{\text{shift}}$ and the $\lambda_j$ are constants. Let $|\{n_j\}\rangle_b $ be eigenstates of the number
operators in the $b$-basis:
\begin{equation}
b^\dagger_j b_j |\{n_j\}\rangle_b = n_j |\{n_j\}\rangle_b \,.
\end{equation}
These states are also the eigenstates of the operator $T$ with eigenvalues
\begin{equation}
\Lambda(\{n_j\}) = \sum\limits_j \lambda_j \, n_j + C_{\text{shift}} \, .
\end{equation}
Note that if the $\lambda_j \geq 0$, then $C_{\text{shift}}$ is the lowest eigenvalue of $T$, and hence is the quantum inequality bound, and the lower limit of the
probability distribution.

In a measurement in the physical vacuum, $|0\rangle_a$, the probability of finding $\Lambda(\{n_j\})$ is the squared overlap of  the eigenstate $|\{n_j\}\rangle_b$
with  $|0\rangle_a$,
\begin{equation}
P(\Lambda(\{n_j\}) =   |_a\langle 0|\{n_j\}\rangle_b|^2 \,.
\end{equation}
Note that the $b$-vacuum, $|0\rangle_b$,  is a squeezed vacuum state in the $a$-Fock space, which is a superposition of all possible even numbers of $a$ particles. Because the 
Bogolubov transformation, Eq.~(\ref{eq:Bogo1}), and its inverse are linear, the one $b$-particle states, $b^\dagger_j |0\rangle_b$, contain only odd numbers of $a$-particles.
This applies to all odd $b$-number eigenstates. This implies that we need only consider eigenstates of $T$ where the $n_j$ are even, as other eigenstates will have vanishing
probability to be found in the physical vacuum, $|0\rangle_a$ . 

It is convenient to divide the set of states $|\{n_j\}\rangle_b$ into different particle number sectors, where the $n$ particle sector
is all states with $n$ particles in various modes, so $n = \sum_j n_j$. Thus the vacuum sector contains only  $|0\rangle_b$, the two particle sector contains states of the form
$|2_j \rangle_b$, two particles in a single mode, or $|1_j, 1_k \rangle_b$, one particle in each of two different modes. The four and higher sectors are constructed similarly.
The numerical computations in Ref.~\citen{SFF18} used $N = 120$ for the $\alpha = 1/2$ case, and included states through the four particle and part of the six particle
sectors.  However, it was found that the dominant
contributions to  $P(x)$ comes from the vacuum and  two particle sector. This can be measured by the cumulative distribution function, $1- P_>(x)$, which should approach
one for large $x$, but will always be slightly less than one in an approximate calculation. It shown in Table 1 of  Ref.~\citen{SFF18} that the four particle sector gives a contribution
of about $1.9 \times 10^{-4}$  when $\alpha = 1/2$, and that of the six particle sector is about two orders of magnitude smaller. 
In Ref.~\citen{WSF21}, $N = 600$ and only the vacuum and  two particle sectors were included. 

\begin{figure}[b]
\centerline{\includegraphics[width=12cm]{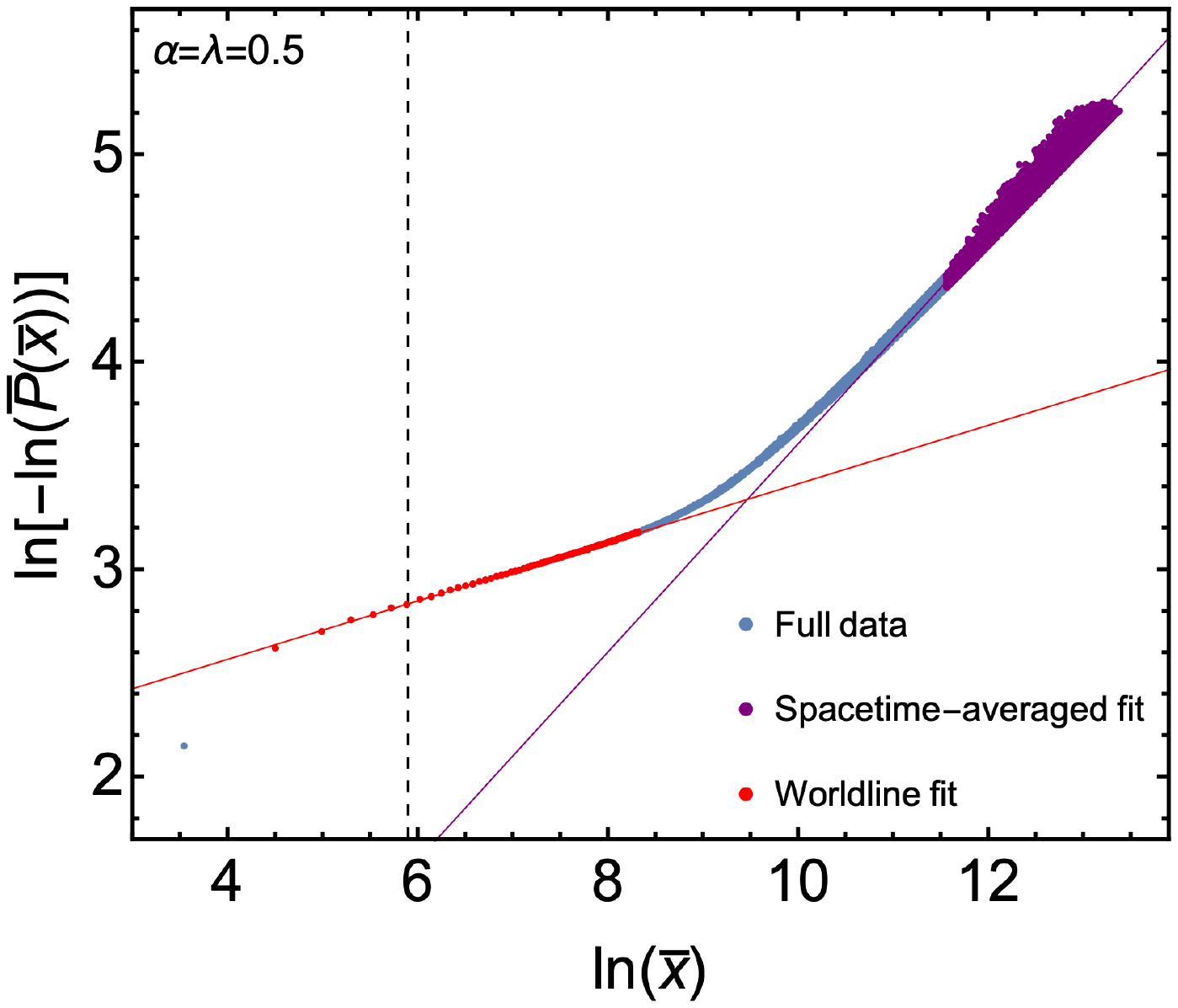}}
\caption{The probability distribution for the space and time averaged $\dot{\varphi}^2$  of a massless scalar field in four dimensional Minkowski spacetime.}      
\label{fig:4D}
\end{figure}

A typical result for the numerical probability distribution is illustrated in Fig.~\ref{fig:4D}\footnote{Originally published as the upper panel in Fig. 2 of Ref.~\citen{WSF21}}.      
Here $\dot{\varphi}^2$  for a massless scalar field has been averaged in
time by a sampling function  with $\alpha = 1/2$ and $\tau = 1$ and in space by a spherically symmetric function with  $\lambda = 1/2$ and $\ell = 0.14$. For 
smaller values of $x$, the plot of $\ln[-\ln(P)]$ as a function of $\ln x$ is  straight line with slope of about $0.14$, which is roughly equal to the predicted value 
of $c=1/6$ for the worldline regime. For larger values of $x$, there is a smooth transition to linear behavior with a slope of $0.501$, which is very close to the
predicted value of $c = 1/2$ for spacetime averaged behavior. The better fit to the predicted value of $c$ in the spacetime averaged regime is probably due to
having more data in that region. In this case, the predicted value of the transition from Eq.~(\ref{eq:x*}) is at $ \ln x_* \approx 6$, which is illustrated by the vertical
dashed line. The actual transition occurs at a somewhat larger value, in the range $8 <  \ln x_* < 10$. The derivation of Eq.~(\ref{eq:x*}) in Sec.~VIB of
Ref.~\citen{FF2020} offers a criterion for the upper limit of validity of the worldline approximation. The numerical results suggest that this criterion may be too
conservative, and the worldline approximation may remain valid to somewhat larger values of $x$.

Overall, these numerical results confirm the behavior of the probability distribution for space and time averaged quadratic operator found by the moments
approach in Ref.~\citen{FF2020}. There is a clear transition from the  worldline regime, where $P(x)$ falls as an exponential of a fractional power, to a
spacetime averaged regime, where it falls as an exponential of a larger fractional power, which reflect the effects of space averaging in addition to time
averaging.

\section{Physical Effects of the Fluctuations}
\label{sec:phy-effect}
In this section we discuss several effects of quantum stress tensor fluctuations and related phenomena.

\subsection{Gravitational and Cosmological Effects}
\label{sec:grav-cosmo}

In Einstein's general theory of relativity, the stress tensor $T_{\mu\nu}$ is the source of gravity through the Einstein equations 
\begin{equation}
G_{\mu\nu} = R_{\mu\nu} - \frac{1}{2} g_{\mu\nu}\, R = 8\pi\, G\,T_{\mu\nu}\,.
\label{eq:Einstein}
\end{equation} 
Here $g_{\mu\nu}$ is the spacetime metric, $G_{\mu\nu}$ and $R_{\mu\nu}$ are the Einstein and Ricci tensors which describe the spacetime
curvature, $R = R^\mu_\mu$ is the scalar curvature, and $G$ is Newton's constant. It is apparent that quantum fluctuations of $T_{\mu\nu}$ 
will drive passive fluctuations of $G_{\mu\nu}$, and hence of the spacetime geometry. These fluctuations are distinct from the active fluctuations
of gravity which will arise if gravity is quantized. Both types of fluctuations are expected in a quantum theory of gravity, so the study of passive
fluctuations can provide some insight into quantum effects in gravity.

\subsubsection{Spacetime geometry fluctuations and light propagation}
\label{sec:light}

Light propagating through a fluctuating spacetime can undergo several related effects, which may have the potential to leave observable signatures.
An example is lightcone fluctuations; the fixed lightcone of classical gravity can undergo quantum fluctuations. Because gravity is non-dispersive, this
effect applies to all light frequencies equally. In principle, lightcone fluctuations lead to a variation in the flight times of pulses, with some traveling slower 
than the mean light speed, and others faster~\cite{F95,FS96}. Some closely related effects were discussed by Hu and Shiokawa~\cite{HS98}.

Both passive and active fluctuations of gravity will modify light propagation, and both can lead to fluctuations of the Riemann tensor and of geodesic
deviation~\cite{VFB18}, which in turn can produce  spectral line broadening~\cite{TF06}.  
However,  there is one effect which is especially sensitive to passive geometry fluctuations,
and hence to quantum stress tensor fluctuations. This is fluctuations  in focussing of light rays~\cite{Moffat,BF04}, which can be studied using the
Raychaudhuri equation: (See, for example, Wald~\cite{Wald}.)
\begin{equation}
\frac{d \theta}{d \lambda} = - R_{\mu\nu} k^\mu k^\nu - \frac{1}{2}\, \theta^2
-\sigma_{\mu\nu} \sigma^{\mu\nu} + \omega_{\mu\nu}  \omega^{\mu\nu} \,.
                                                 \label{eq:ray}
\end{equation}
Here  $\theta$ is the expansion of a bundle of light rays, the logarithmic derivative of the bundle's cross sectional area as a function of the affine parameter,
$\lambda$, which increases along the bundle's propagation direction. Here $k^\mu$ is the null vector tangent to the bundle's worldline, and may be taken to 
be the four-momentum vector of the photons in the bundle. The tensors  $\sigma^{\mu\nu}$ and $\omega^{\mu\nu}$ are the shear and vorticity, and measure
the tendency of the rays in the bundle to either shear or twist relative to one another. In the limit of weak gravitational fields, the quadratic terms in $\theta$,
$\sigma^{\mu\nu}$ and $\omega^{\mu\nu}$ can be small compared to the Ricci tensor term. This term is determined by Eq.~(\ref{eq:Einstein}) in terms
of the stress tensor $T_{\mu\nu}$. In this case, Eq.~(\ref{eq:ray}) becomes a Langevin-type equation in which fluctuations of the stress tensor determine those
of the expansion $\theta$. This  can lead to the blurring of images by  the fluctuating gravitational field~\cite{BF04}.

\subsubsection{Perturbations in Inflationary Cosmology}
\label{sec:inflation2}

As was briefly discussed in Sec.~\ref{sec:inflation}, inflationary models explain the large scale structure of the present day universe as arising from the vacuum
fluctuations of an approximately linear field. This results in a nearly Gaussian probability distribution for the perturbations.  However, there is a possibility that
quantum stress tensor fluctuations could contribute either to the primordial density perturbations~\cite{WKF07,Ford:2010wd}, or to primordial gravitational 
waves~\cite{Wu:2011gk}, and introduce non-Gaussian behavior. These references concentrate on the calculation of the variance of the fluctuations, and there is 
more work to be done to better understand the  spacetime averaging and the probability distributions.

\subsubsection{Effects on the Small Scale Structure of Spacetime}
\label{sec:small-scale}

Carlip, Mosna, and Pitelli~\cite{CMP11,CMP18} have proposed a model in which large stress tensor fluctuations dramatically alter spacetime structure on scales
slightly larger than the Planck length. These authors use a two dimensional model of the form of that discussed in Sec.~\ref{sec:2D} in which stress tensor fluctuations 
lead to light cones closing on a length scale about one order of magnitude larger than the Planck length, and  spacetime breaking into causally
disconnected regions at this scale.

\subsection{Effects in Nonlinear Optics}
\label{sec:nonlinear}

In a nonlinear optical material, the electric polarization vector, and hence the speed of propagation of light pulses can depend upon an applied electric field. This dependence
can be linear (nonzero second order susceptibility), quadratic ( nonzero third order susceptibility), or higher order. If the electric field undergoes quantum fluctuations, this leads
to fluctuations in the speed of pulses~\cite{FDMS13,BDF14,BDFR16}. This is an analog model for the quantum lightcone fluctuations discussed in Sec.~\ref{sec:light},
and is a potentially observable effect in nonlinear optics. These fluctuations may have either a Gaussian probability distribution for linear fields, or a non-Gaussian  distribution
analogous to those for quantum stress tensors treated in Sec.~\ref{sec:pdf}. In the latter case, there can be large fluctuations which depend upon the switching of the
probe light pulse. This might provide an experimental means to study large vacuum fluctuations.

\subsection{Vacuum Radiation Pressure Fluctuations on Charged Particles}
\label{sec:RadPress-charges}

The energy or linear momentum flux in the electromagnetic field is given by the Poynting vector, $\mathbf{ E  \times B}$. When the field is quantized, this becomes
an operator, which has a zero expectation value in the vacuum and will fluctuate  symmetrically around this mean value. This leads to vacuum radiation pressure
fluctuations on a mirror~\cite{WF01} or a charged particle~\cite{HF17}. Here we focus on the latter case. First consider a classical plane electromagnetic wave with
angular frequency $\omega$ which scatters from a  particle with rest mass $m$ and electric charge $q$ by Thompson scattering, for which the total cross section is
\begin{equation}
\sigma_T = \frac{q^4}{6 \pi \, m^2}\,.
\label{eq:Thompson}
\end{equation}
This cross section applies when $\omega \ll m$ and arises from a dipole pattern of scattered radiation. Here the scattered radiation carries no net momentum. The linear 
momentum of the incident radiation is converted to mechanical momentum of the particle, which experiences  a force of  $ \sigma_T \,  \mathbf{ E  \times B}$.
In the case of a quantized electromagnetic field in the vacuum state, the radiation pressure fluctuations lead to force fluctuations and to linear momentum fluctuations
of the particle. The probability distribution for large vacuum radiation pressure fluctuations on a charged particle will have the same asymptotic form  
as that discussed in Sec.~\ref{sec:moments-STA} for operators such as the electromagnetic energy density.  

Radiation pressure fluctuations can enhance the rate at which charged particles penetrate a potential barrier~\cite{HF17}, analogous to the effect of electric field 
fluctuations~\cite{FZ99,Huang:2015lea}. However, now the effect is potentially larger, due to the relatively high probability of large fluctuations. Here the magnitude of the effect
depends sensitively on how the operator $ \sigma_T \,  \mathbf{ E  \times B}$ is averaged. In Ref.~\citen{HF17}, it was suggested that this averaging process might
be determined by a combinations of shape of the incident particle's wavepacket and the shape of the potential barrier. The radiation pressure fluctuations seem to become
more important at higher energies, and might be observable in nuclear fusion reactions.

\subsection{Vacuum Radiation Pressure Fluctuations on Polarizable Particles}
\label{sec:RadPress-atoms}

Vacuum radiation pressure fluctuations on an electrically polarizable particle, such as an atom, take a very different form as compared to the case of 
electric charges~\cite{HF17,F21}. This is because
the dominant form of scattering is now Rayleigh scattering, rather than Thompson scattering. The total cross section for Rayleigh scattering is
\begin{equation}
\sigma_R = \frac{\alpha_0^2}{6 \, \pi} \, \omega^4 \,.
\label{eq:Rayleigh}
\end{equation}
Here $\alpha_0$ is the static polarizability, and $\omega$ is the angular frequency of the incident light, which is assumed to be small compared to typical atomic energy level separations.
The $\omega$-dependence of $\sigma_R$ can be interpreted as leading to additional time derivatives in the quantum operator  describing vacuum radiation pressure fluctuations on an
atom. Take this operator to be
\begin{equation}
 \frac{\alpha_0^2}{6 \, \pi} \;  (\mathbf{\ddot{E}} \times  \mathbf{\ddot{B}}) \,.
  \label{eq:Rayleigh-force}
\end{equation}
In the case that the electric and magnetic field has sinusoidal time dependence, the time derivatives in Eq.~(\ref{eq:Rayleigh-force}) produce the factor of $\omega^4$ in
Eq.~(\ref{eq:Rayleigh}).

The  additional time derivatives have a dramatic effect on the probability of large radiation pressure fluctuations on an atom. Let the dimensionless variable describing these fluctuations
be 
\begin{equation}
x = \tau^8\, |\mathbf{\ddot{E}} \times  \mathbf{\ddot{B}}| \,.
\end{equation}
In the worldline regime, now given by  
\begin{equation}
x < x_* = \left(\frac{\tau}{\ell}\right)^7\,,
\label{eq:x*-atom}
\end{equation}
 the probability distribution has the approximate form of Eq.~(\ref{eq: asym-P}) with $c = \alpha/7$. If the Fourier transform parameter is $\alpha = 1/2$, then the probability
 distribution falls very slowly, with $c =1/14$.
 
 This distribution was used in Ref.~\citen{F21} to estimate vacuum radiation pressure fluctuations on a Rydberg atom, which is an atom in a highly excited state with a very
 large polarizability. Such an atom can be formed by the effect of a switched laser pulse. In the model of Ref.~\citen{F21}, the Fourier transform of this pulse determines the
 parameter $\alpha$ and hence the  probability distribution of the vacuum radiation pressure fluctuations.

 \subsection{Light Scattering from Quantum Fluid Density Fluctuations}
 \label{sec:fluid}
 
 A different analog model for quantum stress tensor fluctuations is provided by quantum fluctuations in the density of a fluid due to quantized sound waves. Several authors
 have discussed the fact that fluid density may be described by an quantum operator which is proportional to $\dot{\varphi}$, where $\varphi$ is a massless scalar field,
 but with the speed of light replaced by the speed of sound~\cite{LP69,Unruh,FF04}. The possibility of observing these density fluctuations by light scattering was discussed
 in Ref.~\citen{FS09}. Here a scattering cross section was treated, which effectively measures the fluid density fluctuations. More recently, a model
 was proposed~\cite{WF20} in which the probability distribution for large fluctuations  might be observed. 
 
\begin{figure}[b]
\centerline{\includegraphics[width=12cm]{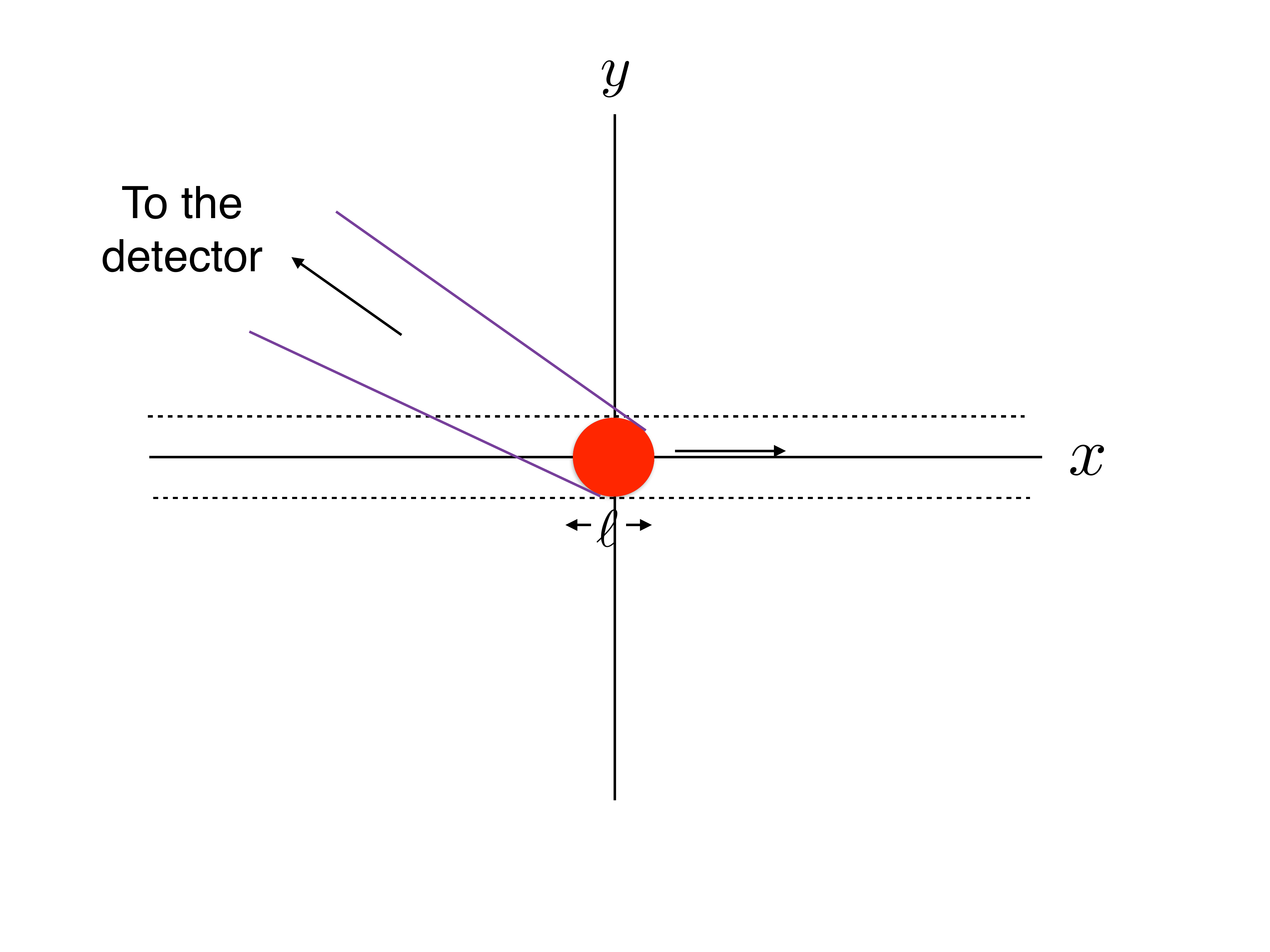}}
\caption{The scattering of light by a large density fluctuation is illustrated. The pulse of light moves to the right at a speed much greater than the speed of sound.
The fluctuation region of spatial size $\ell$ is shown in color, and photons scattered from this region are seen by a detector.}      
\label{fig:fluid-fluct}
\end{figure}

 The basic idea is to send pulses of light described by compactly supported functions through the fluids, as illustrated in 
 Fig.~\ref{fig:fluid-fluct}\footnote{Originally published as Fig. 2 in.Ref.~\citen{WF20}.}
Because the light scattering is a coherent process, the expected number of photons scattered by a density fluctuation is proportional to the square of the density variation
and hence proportional to $\dot{\varphi}^2$. 
The mean number of photons scattered, averaged over trials, is  measure of the variance of $\dot{\varphi}^2$. However, a histogram of
numbers of scattered photons effectively generates a plot of the probability distribution of the space and time averaged $\dot{\varphi}^2$  operator.

\section{False Vacuum Decay in Field Theory}
\label{sec:falsevac}

A self-coupled scalar field $\varphi$ can be described by a potential $V(\varphi)$. Suppose that the potential has a local minimum at $\varphi = \varphi_0$, which is not a global
minimum.  This minimum is stable against small classical perturbations, but is expected to be unstable in quantum field theory. This is the analog of a quantum particle in a local
potential minimum, which is unstable against quantum tunneling. In field theory, such an unstable configuration is called a false vacuum and is unstable against quantum tunneling
to the global minimum, the true vacuum. A theory which describes this process was developed by Coleman~\cite{C77} and by Callen and Coleman~\cite{CC77}. In this approach,
the tunneling amplitude is computed in euclidean space from a path integral, and the dominant contribution is assumed to come a configuration with minimum euclidean action,
called a ``bounce" or ``instanton". This leads to a decay rate per unit volume proportional to $\exp(-B)$, where $B$ is the  euclidean action of the bounce. This approach is
analogous to the WKB method for computing tunneling rates in quantum mechanics. 

Just as quantum tunneling rates of charged particles can be modified by vacuum electric field or radiation pressure fluctuations, we can ask if quantum field fluctuations can
enhance the rate of false vacuum decay. The effect of linear field fluctuations of $\varphi$ or its time derivative $\dot{\varphi}$ has been discussed by several 
authors~\cite{Linde92,CV99,CRV02,CRV01,ACRV03,CV06, BJPPW19,HY19,BDV19,Wang19,HF22}. These authors generally agree that linear field fluctuations give a
contribution to the decay rate which is of the same order as the $\exp(-B)$ result in Ref.~\citen{C77}. However, there is disagreement as to whether field fluctuations
represent a physically distinct decay channel. The order of magnitude agreement can be viewed as arising from the Gaussian probability distribution, Eq.~(\ref{eq:gaussian}),
for vacuum fluctuations of linear fields, which has a similar functional form to $\exp(-B)$. A heuristic picture of the effect of a large $\dot{\varphi}$ fluctuations was given in
Ref.~\citen{HF22}, in which this effect is analogous to a large initial value of $\dot{\varphi}$ for a classical field in a local potential minimum. This can cause a finite region to
evolve classically over the barrier between the local minimum and the global minimum. 

The effects of quadratic operator fluctuations, such as those of $\dot{\varphi}^2$, were also discussed in  Ref.~\citen{HF22}, where it was argued that these can have a larger
effect than either quantum tunneling or linear field fluctuations, This is due to the  slower than exponential decay of the probability distribution for $\dot{\varphi}^2$, and suggests
that these may be the dominant contribution in some cases.

\section{Summary}
\label{sec:final}

In this article, various aspects of quantum field fluctuations have been reviewed, beginning with vacuum fluctuations of the electric field. These play a role in the Lamb shift
and can enhance quantum tunneling of charged particles. Similar fluctuations of a scalar inflaton field could have been the source of the primordial density fluctuations in the early universe
which gave rise to galaxies. The Gaussian probability distribution of these fluctuations was discussed, and it was argued that a meaningful discussion of the fluctuations
requires that the field be averaged in time or in space, and that this averaging describes the measurement process. Furthermore, the averaging should be over finite regions,
requiring averaging functions with compact support, those which strictly vanish outside of a finite region and are hence non-analytic. 

Such functions are also essential for the treatment of the fluctuations of quadratic operators, such as energy density or flux. Here the probability distribution becomes very
sensitive to the details of the averaging and hence of the measurement process. The probability distribution for vacuum fluctuations can now fall relatively slowly, as an exponential
of a fractional power. This leads to the possibility of large fluctuations and a dominance of vacuum fluctuations over thermal fluctuations. A quadratic operator averaged in
both space and time typically has a probability distribution exhibiting two distinct regimes. The first is a worldline regime describing smaller magnitude fluctuations and which
depends primarily upon the temporal averaging. Larger magnitude fluctuations are in a spacetime averaged regime, where both time and space averaging are important. Here the
probability distribution falls faster than in the worldline regime, but still as an exponential of a fractional power. This general behavior is supported both by analytic arguments
on the rate of growth of the moments of the distribution, and by a numerical diagonalization approach. 

Large stress tensor fluctuations can induce passive fluctuations of the gravitational field, a variety  of quantum gravity effect, and in principle produce quantum lightcone fluctuations,
and other effects on light rays which could play a role in the early universe. There are analog systems, such as nonlinear optical materials or fluid zero point density fluctuations
which have similar effects to those in gravity, and which might be experimentaly accessible. This is an ongoing area of research.

\section*{Acknowledgments}
I would like to thank Haiyun Huang, Enrico Schiappacasse, and Peter Wu for useful conversations.
This work was supported in part  by the National Science Foundation under Grant PHY-1912545.

\end{document}